\documentstyle[aps]{revtex}

\draft

\def\bfr{{\bf r}}

\def\bfr{{\bf r}}

\def\tpsi{\tilde \psi (\bfr)}
\def\tpsid{\tilde \psi^\dagger (\bfr)}

\begin{document}

\title{Gapless finite-$T$ theory of collective modes of a trapped gas}
 
\author{D.\ A.\ W.\ Hutchinson, R.\ J.\ Dodd\cite{NIST}, and K.\ Burnett\cite{NIST}}
 
\address{Clarendon Laboratory, Department of Physics, University of Oxford,\\
Parks Road, Oxford OX1 3PU}
 
\date{\today}
 
\maketitle
 
\abstract{
We present predictions for the frequencies of collective modes of
trapped Bose-condensed $^{87}$Rb atoms at finite temperature. Our
treatment includes a self-consistent treatment of the effects upon
the mean-field from
finite-$T$ excitations and the anomalous average. This is the first
gapless \cite{GRIFFIN} 
calculation of this type for a trapped Bose-Einstein condensed gas.
The corrections 
quantitatively account for the downward shift in the $m=2$ excitation
frequencies observed in recent experiments 
as the critical temperature is approached}
 
\pacs{PACS Numbers: 03.75.Fi, 05.30.Jp, 67.40.Db}

%%%%%%%%%%%%%%%%%%%%%%%%%%%%%%%%%%%%%%%%%%%%%%%%%%%%%%%%%%%%%%%%%%%%%%
%%%%%%%%%%%%%%%%%%%%%%%%%%%%%%%%%%%%%%%%%%%%%%%%%%%%%%%%%%%%%%%%%%%%%%
%%%%%%%%%%%%%%%%%%%%%%%%%%%%%%%%%%%%%%%%%%%%%%%%%%%%%%%%%%%%%%%%%%%%%%

Measurements of the collective excitation frequencies of trapped,
Bose-Einstein condensed alkali vapours are providing the most stringent
tests of the theoretical understanding of these newly produced systems.
Mean-field theories 
have been used, with great success, both qualitatively and
quantitatively in determining the excitation frequencies of the
condensates, especially at relatively low temperatures ($\leq$ 0.7
$T_c$) \cite{HUTCHINSON}. These calculations have been based upon the
Popov approximation to Hartree-Fock Bogoliubov (HFB) theory, where the
anomalous average of the fluctuating field operator is neglected, or
upon simpler versions of (finite-$T$) mean-field 
theory \cite{stringari} . However,
recent experimental results from JILA\cite{CORNELL} indicate a
discrepancy with these theoretical results as one approaches the
critical temperature. This has raised questions about the validity of
the HFB-Popov approach and mean fielf theories in general,
when the condensate is strongly depleted and
lead to debate as to how one can {\it consistently} improve upon this
theory. In particular, if one retains a contact ($\delta$-function)
interaction potential and simply includes the self-consistent
anomalous average, a gap is found in the excitation
spectrum\cite{GRIFFIN}. In addition, the most naive treatment of the
anomalous average results in a (ultra-violet) divergent quantity if
one retains a simple contact interaction. The
validity of replacing the true interaction with a contact interaction
must be addressed at the same time as the divergence if one is to go
beyond the Popov approximation in a consistent manner.

In this letter we address these questions, with a consistent,
gapless theory that goes beyond the Popov approximation.
We show how the anomalous average should be renormalised in a 
full computation to remove
the resulting ultra-violet divergence, in a manner that is closely 
related to
those developed using a many-body T-matrix
approach for homogeneous systems \cite{MORGAN,SHI}. 
We then examine what changes these
extensions make to the predictions of the theory in both an isotropic
geometry and for a trap corresponding to the JILA experiments. Our
results are then compared in detail to those obtained experimentally.

%%%%%%%%%%%%%%%%%%%%%%%%%%%%%%%%%%%%%%%%%%%%%%%%%%%%%%%%%%%%%%%%%%%%%%
%%%%%%%%%%%%%%%%%%%%%%%%%%%%%%%%%%%%%%%%%%%%%%%%%%%%%%%%%%%%%%%%%%%%%%
%%%%%%%%%%%%%%%%%%%%%%%%%%%%%%%%%%%%%%%%%%%%%%%%%%%%%%%%%%%%%%%%%%%%%%

We start  with a
brief review of the HFB theory. The treatment
presented here follows closely that previously discussed by Griffin
\cite{GRIFFIN} and yields the collective excitations of the {\it condensate}
in the presence of a {\it static} thermal cloud as in previous studies.
The condensate wavefunction, $\Phi(\bfr)$, is obtained from
the generalised Gross-Pitaevskii equation \cite{GRIFFIN}
\begin{equation}
\left\{ -{\nabla^2 \over 2M} + V_{ext}(\bfr) + g[n_c(\bfr) + \tilde m(\bfr) +
2\tilde
n(\bfr) ] \right\}
\Phi(\bfr) = \mu \Phi(\bfr).
\end{equation}
Here we have made the usual decomposition of the Bose field operator,
$\hat{\psi}(\bfr)$, into condensate and noncondensate parts, i.e.,
$\hat{\psi}(\bfr) =
\Phi(\bfr)+\tilde{\psi}(\bfr)$. The terms involving the interaction
strength, $g=4 \pi \hbar ^2 a/M$, arise from the use of a contact interaction,
$g \delta (\bfr)$, where $a$ is the scattering length measured for binary
scattering in vacuo. 

The collective excitations are then given by 
the coupled HFB equations \cite{GRIFFIN} 
\begin{eqnarray}
\hat {\cal L} u_i(\bfr) &-& g [n_c(\bfr) + \tilde m(\bfr)] v_i(\bfr) = E_i
u_i(\bfr) \\
\hat {\cal L} v_i(\bfr) &-& g [n_c(\bfr) + \tilde m(\bfr)] u_i(\bfr) = -E_i
v_i(\bfr),
\end{eqnarray}
with 
\begin{equation}
\hat {\cal L} \equiv -\frac{\nabla^2}{2M} + V_{ext}(\bfr)+ 2gn(\bfr) - \mu
\equiv \hat h_0 + g[n_c(\bfr)- \tilde m(\bfr)],
\end{equation}
defining the quasiparticle excitation energies $E_i$ and
amplitudes $u_i$ and $v_i$. Here $n_c(\bfr) \equiv |\Phi(\bfr)|^2$ is the
density of condensed atoms,
$\tilde n(\bfr) \equiv \langle \tpsid \tpsi \rangle$ gives the excited state
population density and $\tilde
m(\bfr) \equiv \langle \tpsi \tpsi \rangle$ is the anomalous average.

The expressions for $\tilde{n}(\bfr)$ and $\tilde{m}(\bfr)$ 
in terms of the quasiparticle spectrum are;
\begin{equation}
\tilde n(\bfr) = \sum_i \left \{ |v_i(\bfr)|^2 +
\left [ |u_i(\bfr)|^2+|v_i(\bfr)|^2 \right ] N(E_i)
\right \},
\end{equation}
and
\begin{equation}
\tilde m(\bfr) = -\sum_i u_i^*(\bfr)v_i(\bfr)\left \{2N(E_i)+1
\right \},
\end{equation}
where the Bose factor is given by
\begin{equation}
N(E_i)=\frac{1}{e^{\beta E_i}-1}.
\end{equation}
The HFB equations provide a variationally lowest free-energy for
the system and
could, in principle, be used as they stand. They cannot however
guarantee to give the best excitation frequencies. Indeed it is
well know \cite{girardeau} that the inclusion of the anomalous
average leads to a theory with a (unphysical) gap \cite{GRIFFIN}
in the excitation
spectrum. This can be seen to arise from the fact that the effective
interaction between a pair of particles depends upon whether both
come from the condensate or one is excited.
The standard treatment in calculations for trapped gases has
been to {\it neglect} $\tilde{m}(\bfr)$ in the above equations, which
restores the symmetry and hence
leads to a gapless theory. This is the Popov approximation.

In addition, one finds that the anomalous average is divergent if
one uses an bare contact interaction.
To go beyond Popov one has to renormalise the anomalous
average to remove this ultra-violet divergence. This is done by
noting that $\tilde m$ effectively alters the interaction strength of
the particles due to the presence of the condensate.  The interaction
strength used is based upon measurements of the scattering length {\it
in vacuo}. One should, therefore, subtract off the vacuum perturbative limit of
$\tilde m$, as these effects are already
included in the measured scattering length \cite{lifshitz}. 
At high
energies the perturbative and HFB values for $\tilde m$ are
equivalent, hence the ultra-violet divergence is removed by the
subtraction.
A
second, simpler, renormalisation involves subtracting the ``zero-$T$''
component of $\tilde{m}(\bfr)$ from itself (i.e. dropping the $1$ in
the $\{2N(E_i)+1\}$ term). This can be seen, in a homogeneous system,
to be equivalent to the {\em correct} renormalisation, e.g.,
\begin{eqnarray}
\lim_{k\rightarrow\infty} \tilde{m}(T) &=& \lim_{k\rightarrow\infty} 
\int d{\bf k} u_{\bf k} v_{\bf k} \left\{ 2  N(E_k)+1 \right\} \sim 
\lim_{k\rightarrow\infty} \int d{\bf k} u_{\bf k} v_{\bf k} \{ 1\}.
\end{eqnarray}
Relaxing the requirement of taking the limit makes this
renormalisation equivalent to dropping the $1$ in the $\{2N(E_i)+1\}$
term.
 
This procedure renders the HFB theory non-divergent, but leaves it with
a gap in the excitation spectrum. This gap is due,
as we remarked above, to the
inconsistent treatment of interactions between particles \cite{GRIFFIN}.
One would expect the effective interactions between any pair of particles,
whether both come from the condensate or otherwise, to be the same. In HFB this 
is not the case. To be precise, one should expect there to be a dependence
on the relative momentum of the pair, but this will be weak and we shall
ignore it (as discussed in \cite{MORGAN} and \cite{STOOF}).

To go beyond Popov {\it consistently} one has to treat the
inter-particle interactions in a different manner. One can retain the
gapless nature of the HFB-Popov equations which neglect $\tilde m$ in
both the generalised Gross-Pitaevskii equation and in the HFB
equations by simply modifying the interaction strength $g$ by making
the substitution \cite{MORGAN}
\begin{equation}
g \longrightarrow g \left \{ 1 + \frac{\tilde m(\bfr)}{n_c(\bfr)} \right \}.
\end{equation}
The form of this equation is motivated by the formal discussions given
by Stoof \cite{STOOF} and Proukakis \cite{NICK} and
is equivalent to the low momentum limit of the many body T-matrix for a
homogeneous system \cite{STOOF,SHI}.

These two steps generate a
self-consistent, gapless, non-divergent theory which goes beyond the
Popov approximation. Indeed this is the {\it first} consistent
step, in terms of the treatment of the inter-particle interactions, that
one can take beyond the Popov approximation to the HFB theory.
In the homogeneous limit this treatment
is related to the Beliaev formalism if the thermal
cloud is taken to be static and any damping effects are ignored.

%%%%%%%%%%%%%%%%%%%%%%%%%%%%%%%%%%%%%%%%%%%%%%%%%%%%%%%%%%%%%%%%%%%%%%
%%%%%%%%%%%%%%%%%%%%%%%%%%%%%%%%%%%%%%%%%%%%%%%%%%%%%%%%%%%%%%%%%%%%%%
%%%%%%%%%%%%%%%%%%%%%%%%%%%%%%%%%%%%%%%%%%%%%%%%%%%%%%%%%%%%%%%%%%%%%%

We now present results from three different formalisms;
Hartree-Fock-Bogoliubov with Popov approximation (Popov),
Hartree-Fock-Bogoliubov (HFB) and the gapless Hartree-Fock-Bogoliubov
(GHFB). Each theory represents a closed set of equations which can be
self-consistently solved numerically.  First we present results from
each of the calculations for an isotropic trap of frequency 200 Hz
containing 2000 rubidium atoms and for an anisotropic trap
corresponding to the recent JILA experiments.

In the case of the isotropic trap with relatively small numbers of
atoms the differences between the different treatments are virtually
unobservable, see Fig.\ \ref{FIG1}. Indeed even when the number of
particles is increased in the isotropic trap, the shifts in the
excitation frequencies are only of the order of 1 \%.

For the anisotropic trap, here chosen to correspond to the
JILA experiment\cite{CORNELL}, there is small but significant change from the
Popov results, cf.\ Fig.\ \ref{FIG2}. Even here the shift is only of
the order of 2-3 \% and is downward in frequency for both the $m=2$
and $m=0$ modes. This shift does now agree {\it quantitatively} 
with the experimental data for
the lower energy excitation, but does not agree with the
stated shifts in the higher energy $m=0$ mode.

%%%%%%%%%%%%%%%%%%%%%%%%%%%%%%%%%%%%%%%%%%%%%%%%%%%%%%%%%%%%%%%%%%%%%%
%%%%%%%%%%%%%%%%%%%%%%%%%%%%%%%%%%%%%%%%%%%%%%%%%%%%%%%%%%%%%%%%%%%%%%
%%%%%%%%%%%%%%%%%%%%%%%%%%%%%%%%%%%%%%%%%%%%%%%%%%%%%%%%%%%%%%%%%%%%%%

In this letter we have shown how to improve consistently upon
the Popov approximation to the HFB treatment of the
collective excitations of a trapped condensate  and 
that the corrections to the
results previously obtained (within the Popov approximation) are
small. These corrections are in quantitative agreement with the
observed downward shift in the frequency of the $m=2$ mode in
the JILA experiment near $T_c$. If the upper mode observed in the
experiment is indeed the $m=0$ mode at all $T$, then an extension to
the theory that treats the
thermal cloud of atoms in a dynamical  manner is going to 
be needed to explain the
{\it upward} shift in the higher energy mode.

%%%%%%%%%%%%%%%%%%%%%%%%%%%%%%%%%%%%%%%%%%%%%%%%%%%%%%%%%%%%%%%%%%%%%%
%%%%%%%%%%%%%%%%%%%%%%%%%%%%%%%%%%%%%%%%%%%%%%%%%%%%%%%%%%%%%%%%%%%%%%
%%%%%%%%%%%%%%%%%%%%%%%%%%%%%%%%%%%%%%%%%%%%%%%%%%%%%%%%%%%%%%%%%%%%%%
\section{Acknowledgements}

We would like to thank S.\ A.\ Morgan and N.\ P.\ Proukakis for
discussion, Charles Clark and Mark Edwards for much assistance and
Alan Griffin for much useful discussion. One of us (DAWH) would like 
to thank Eugene Zaremba in particular,
for his support and assistance over a long period, without which he
could not have contributed to this work.
This work was supported by the UK-EPSRC, US
National Science Foundation Grants No. PHY-9601261 and No. PHY-9612728,
and the US Office of Naval
Research.

%%%%%%%%%%%%%%%%%%%%%%%%%%%%%%%%%%%%%%%%%%%%%%%%%%%%%%%%%%%%%%%%%%%%%%
%%%%%%%%%%%%%%%%%%%%%%%%%%%%%%%%%%%%%%%%%%%%%%%%%%%%%%%%%%%%%%%%%%%%%%
%%%%%%%%%%%%%%%%%%%%%%%%%%%%%%%%%%%%%%%%%%%%%%%%%%%%%%%%%%%%%%%%%%%%%%

%%%%%%%%%%%%%%%%%%%%%%%%%%%%%%%%%%%%%%%%%%%%%%%%%%%%%%%%%%%%%%%%%%%%%%
%%%%%%%%%%%%%%%%%%%%%%%%%%%%%%%%%%%%%%%%%%%%%%%%%%%%%%%%%%%%%%%%%%%%%%
%%%%%%%%%%%%%%%%%%%%%%%%%%%%%%%%%%%%%%%%%%%%%%%%%%%%%%%%%%%%%%%%%%%%%%

%%
%%
\begin{figure}
\caption{The calculated excitation frequencies for 2000 $^{87}$Rb atoms in a 
$200$ Hz spherical harmonic trap. The ideal gas (solid line), GHBF (`$+$'), 
HFB (`$\times$'), and HFB-Popov (`$\circ$') results.}
\label{FIG1}
\end{figure}
\begin{figure}
\caption{The experimental, temperature dependent excitation
spectrum in the JILA TOP trap (filled circles) versus the HFB-Popov
predictions for the $m=0$ mode (top, labelled by ``$+$'') and the $m=2$
mode (bottom, labelled by ``$\times$'') and the GHFB results (`$\circ$').  The
solid curves are excitation frequencies for a {\em zero-temperature}
condensate having the same number of condensate atoms as the
experimental condensate in the finite--$T$ cloud.}
\label{FIG2}
\end{figure}

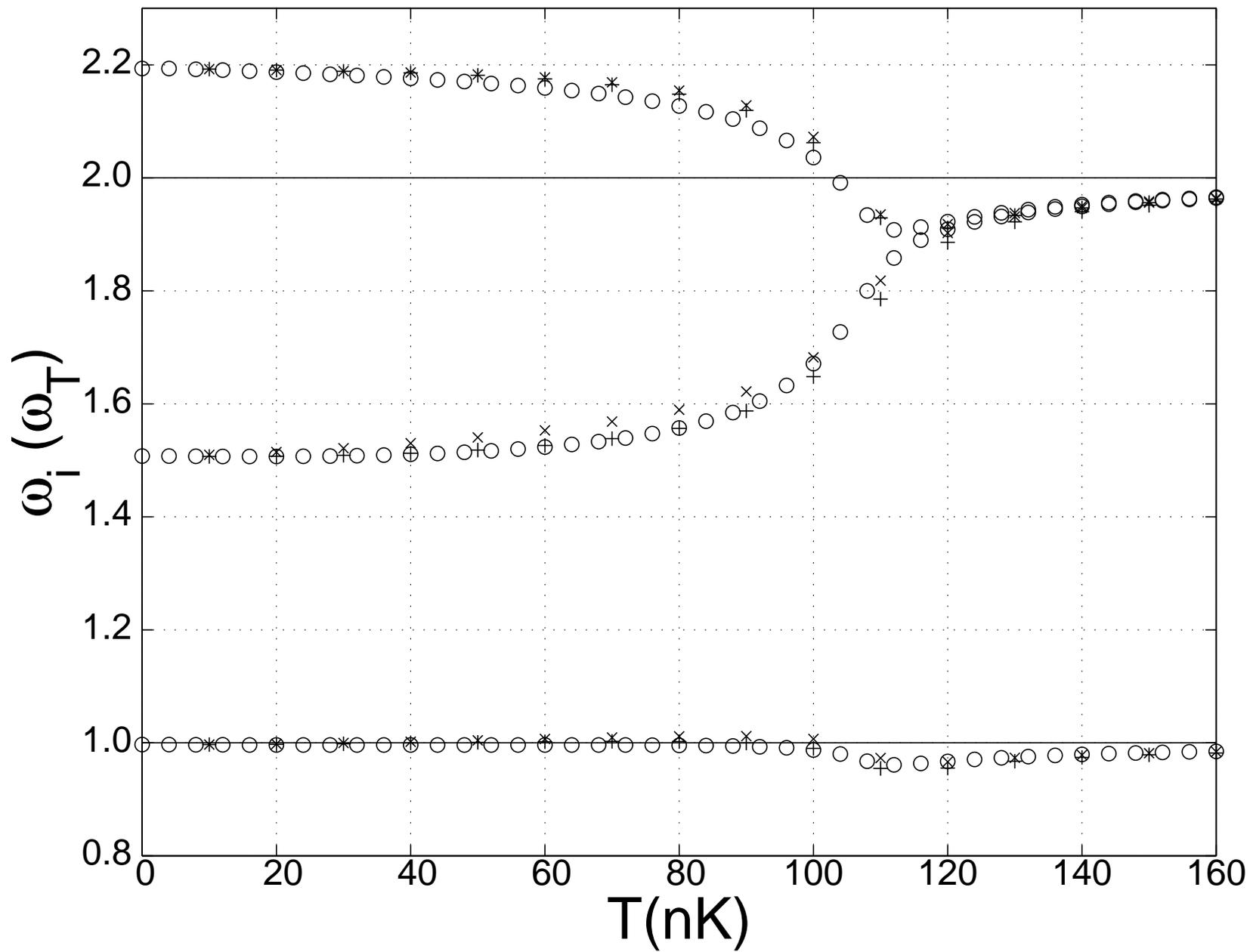
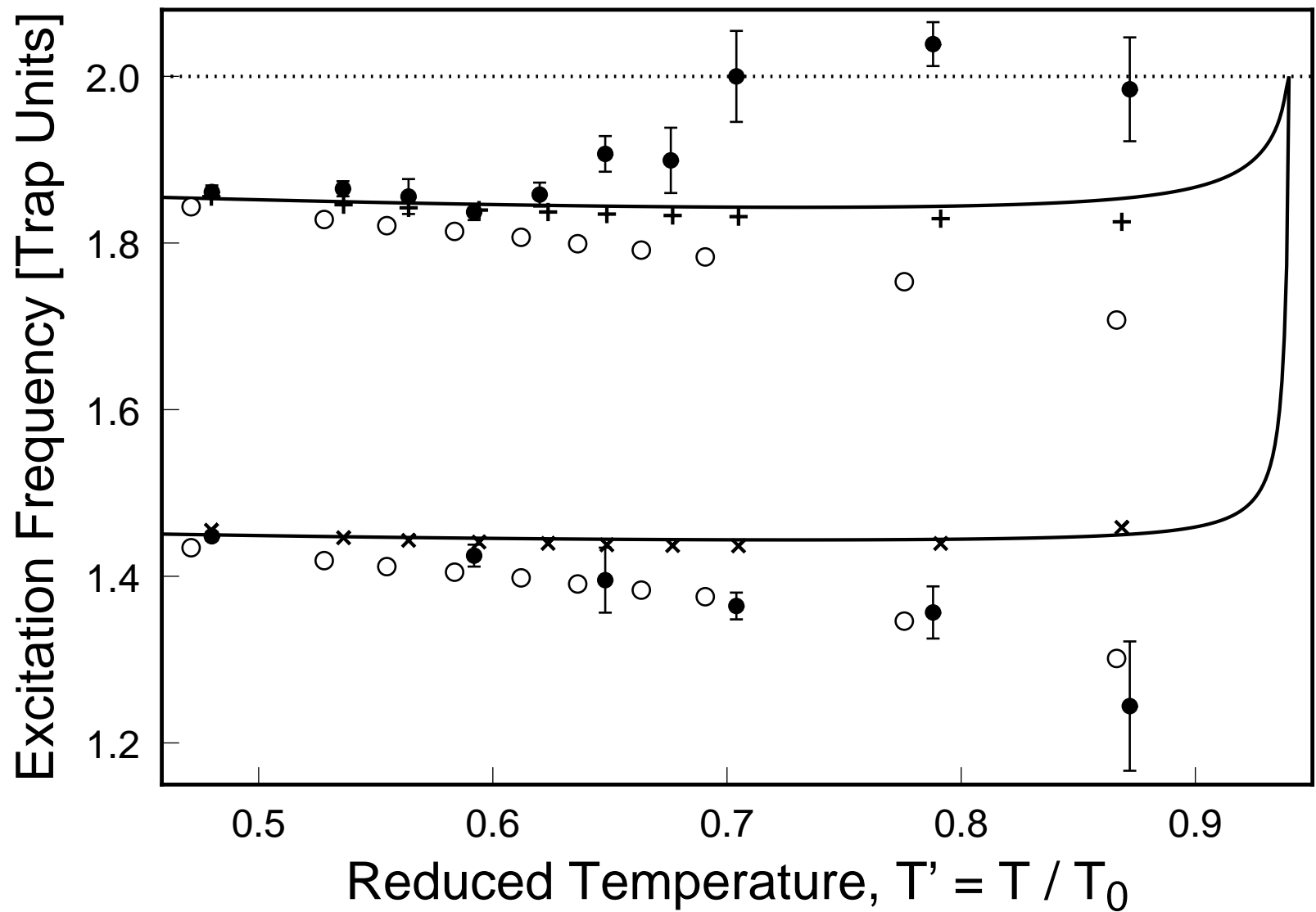
\end{document}